\documentclass[pra,aps,twocolumn,amsmath,amssymb,superscriptaddress]{revtex4-2}
\usepackage{epsfig,amsmath}
\usepackage{subfigure}
\usepackage{graphicx}
\usepackage{dcolumn}
\usepackage{stmaryrd}
\usepackage{mathrsfs}
\usepackage{pifont}
\usepackage{amsthm}
\usepackage{amssymb}
\usepackage{bm}
\usepackage{latexsym}
\usepackage[colorlinks=true,linkcolor=blue,citecolor=blue]{hyperref}
\usepackage{color}
\usepackage{epstopdf}
\usepackage[ruled]{algorithm2e}
\usepackage{lipsum}

\begin{document}

\title{Universal quantum control in biorthogonal framework}

\author{Chen-yi Zhang}
\affiliation{School of Physics, Zhejiang University, Hangzhou 310027, Zhejiang, China}

\author{Jun Jing}
\email{Contact author: jingjun@zju.edu.cn}
\affiliation{School of Physics, Zhejiang University, Hangzhou 310027, Zhejiang, China}

\date{\today}

\begin{abstract}
As a paradigm for describing the open-quantum-system dynamics, the non-Hermitian Schr\"odinger equation generally modifies the state norm, impeding the coherent control over the system. With a general instantaneous representation spanned by biorthogonal ancillary bases, we develop a biorthogonal quantum-control theory for finite-dimensional systems under a general time-dependent non-Hermitian Hamiltonian. In particular, we derive a necessary and sufficient condition for extracting the transitionless pathways that transport parallel in both ket and bra states. When this condition is satisfied by a set of projectors constructed with biorthogonal basis states, the non-Hermitian Hamiltonian becomes diagonal in the relevant biorthogonal subspace, enabling arbitrary initial state to evolve as a superposition of several transitionless passages. Our theory is justified by the bidirectional population transfer, a non-Hermitian holomonic quantum gate implemented in a three-level system, and chiral state conversion. For the bidirectional transfer and the holonomic gate, we can further identify the parametric constraints under which a unitary transformation is formed to link the initial and target states. As a result, the final state is automatically normalized even though the transitional dynamics is nonunitary. In population transfer, controlled gain and loss cut the required coherent pulse area, delivering shorter transfer times than the Hermitian benchmark at equal average Rabi frequency. For the chiral state conversion, we find that a closed parametric loop along opposite directions leads to distinct final state, which is irrelevant to encircling the exceptional point or not. Our work integrates the universal quantum control into the biorthogonal framework, displaying full-rank coherent control in non-Hermitian systems.
\end{abstract}

\maketitle

\section{Introduction}

Precise manipulation of quantum system is fundamental to state preparation~\cite{Law1996PRL,Chou2017Nature,Li2017PRL}, quantum gate design~\cite{Zanardi1999PLA,Schafer2018Nature,Lu2019Nature}, and quantum information processing~\cite{Gisin2007NatPho,Kimble2008Nature,Koch2022EPJ}. The central goal of coherent control is to design time-dependent control fields that steer a quantum system through a prescribed evolution pathway. A wide range of methods have been developed over the past several decades. Adiabatic passage follows instantaneous eigenstates of a slowly varying Hamiltonian~\cite{Kral2007RMP,Vitanov2017RMP}; counterdiabatic driving suppresses transitions among different adiabatic passages~\cite{Berry2009JPA,Odelin2019RMP}; and inverse engineering constructs a Hamiltonian by virtue of dynamical invariants~\cite{Chen2011PRA,Chen2012PRA}, to name a few.

Beyond closed quantum systems, the control over open systems has attracted growing interest in recent years~\cite{Verstraete2009NP,Gautier2025PRL}. As an alternative paradigm with respect to the master equation, the non-Hermitian Schr\"odinger equation provides an effective description of open-system dynamics in certain models~\cite{Minganti2019,Ashida2020}, extending control methods from unitary to nonunitary evolutions. The interplay between coherent operations and non-Hermitian terms yields rich spectral and dynamical phenomena, including exceptional points (EPs)~\cite{Heiss2012JPA,Naghiloo2019NP}, nonreciprocal transport~\cite{Yao2018PRL,Mandal2020PRL}, and direction-dependent state transformation~\cite{Doppler2016Nature,Liu2021PRL}. These features offer extra options to quantum control, but also present challenges for a desired manipulation. In particular, the conventional adiabatic theorem does not generally guarantee eigenstate tracking~\cite{PRA2016Jing} in the open-quantum-system scenario. For non-Hermitian systems, the eigenstate nonorthogonality, the imaginary part of the energy, and the non-conjugate coupling can induce nonadiabatic transitions, even under an extremely slow variation in parameter~\cite{Wang2018PRA}. Controlling non-Hermitian systems requires accounting explicitly for asymmetrical gain and loss in the superposed basis states during the nonunitary dynamics.

Several strategies have been adapted to non-Hermitian systems, including counterdiabatic shortcuts to adiabaticity~\cite{Ibanez2011,Luan2022PRA,Chavva2025PRR,Erdamar2026PRXQ}, dynamical-invariant-based inverse engineering~\cite{Ibanez2011,Fring2022PLA}, and holonomic transformation~\cite{Pinske2019PRA,Sun2026AQT}. Engineered gain and loss can compensate nonadiabatic Hamiltonian, enabling accelerated population transfer in two- and three-level systems~\cite{Torosov2014PRA}. A unified consistency condition with respect to both Hermitian and non-Hermitian systems has been presented by universal quantum control (UQC), which gives rise to nonadiabatic passages in the time-dependent ancillary representation~\cite{JinJing2025Perspective,JinJing2025NonHermitian}. A first attempt leverages the triangularization of the non-Hermitian Hamiltonian in the ancillary picture to activate a single useful passage~\cite{JinJing2025NonHermitian}, which can be end up with an automatically normalized target state under proper constraints.

However, the existing non-Hermitian control protocols can construct only one ad hoc passage in either ket or bra space, without imposing transitionless control across a complete dual set of biorthogonal bases~\cite{Torosov2013PRA,Torosov2014PRA,JinJing2025NonHermitian}. Biorthogonal quantum mechanics provides a formal and consistent framework for understanding the non-Hermitian systems by a key duality relation between their right states in ket space and the associated left partners in bra space and the transformations they can undergo~\cite{Brody2014,Ashida2020}. In particular, a complete set of generally non-orthogonal right states is paired with a set of left states satisfying biorthogonality and a resolution of the identity operator. Probabilities, projectors, and dynamics can then be full-fledged conceptualized in biorthogonal representation, retaining the algebraic structure analog to Hermitian quantum mechanics. Moreover, right and left states evolving under respectively the system Hamiltonian and its adjoint conserve their biorthogonal relation despite their individual standard norms vary with time. This characteristic allows non-orthogonal state geometry and nonunitary evolution to share a common description, providing a natural foundation for quantum control over biorthogonal evolutions.

In this work, we develop a quantum-control framework for non-Hermitian systems via transformation from the time-dependent biorthogonal basis to the stationary biorthogonal basis with rotation operator defined by a complete set of biorthogonal ancillary bases. We prove that the von Neumann equation for each ancillary projector is necessary and sufficient for decoupling the relevant passage from the others. When this condition holds for all projectors of a complete set of biorthogonal basis, the non-Hermitian Hamiltonian becomes diagonal and the resulting controlled evolution operators are full-rank with respect to the system's Hilbert space, equivalent to finding the exact solution of the non-Hermitian dynamics. The controlled evolution is then consisted of a superposition of transitionless passages. Under proper parameters for rotation, the accumulated complex phases of passages determined by both non-Hermitian Hamiltonian and non-Hermitian gauge potential can become real numbers at the end of the passage, leading to an effective unitary transformation and an automatically normalized wavefunction of the system. A bidirectional population transfer can be realized under the same time-dependent Hamiltonian of a non-Hermitian two-level system, whose speed can be faster than its Hermitian counterpart with the same average Rabi frequency by leveraging gain and loss as extra control resources. A unitary holonomic quantum gate can be constructed in a three-level $\Lambda$ system by cancelling the accumulated dynamical phase and the imaginary part of the complex geometric phase over a parametric cycle. When retaining the imaginary part of the passage phase, we can realize the chiral state conversion in a two-level system with and without EP encircling. Either bare-level $|0\rangle$ or $|1\rangle$ can be prepared into one of the two eigenstates, determined by the direction of the parametric variation. And during the control evolution, the occupation on the two eigenstates relies on the direction-dependent amplification and attenuation. Our work substantially formulates the UQC theory~\cite{JinJing2025Perspective,JinJing2025NonHermitian, JinJing2026nonHermitianCV,JinJing2026FromLiouville} in a completed biorthogonal representation, where the biorthogonal ancillary basis gives rise to a symmetrical differential manifold with a non-Hermitian gauge potential.

The rest of this paper is structured as follows. Section~\ref{sec:framework} provides a general theory for engineering multiple nonadiabatic passages in biorthogonal pairs under a time-dependent non-Hermitian Hamiltonian. Section~\ref{sec:popu_transfer} applies the theory to a non-Hermitian two-level system and demonstrates bidirectional population transfer and acceleration by non-Hermitianity. Section~\ref{sec:gate} develops a geometric single-qubit gate in a three-level $\Lambda$ system and numerically verifies its nonunitary dynamics and unitary transformation between two endpoints. Section~\ref{sec:chiral} derives the exact evolution operator for chiral state conversion and demonstrates directional control loops with and without EP encircling. Section~\ref{sec:conclusion} discusses the experimental feasibility of our protocol in simulating non-Hermitian two-level systems, followed by a summary of the main results.

\section{Theoretical framework}\label{sec:framework}

We consider an $N$-dimensional system governed by a general time-dependent non-Hermitian Hamiltonian $H(t)\neq H^\dagger(t)$. Within the biorthogonal framework~\cite{Brody2014,Ashida2020}, the states in the bra and ket spaces satisfy the dual Schr\"odinger equations: ($\hbar=1$)
\begin{subequations}\label{eq:bioshro}
\begin{align}
i\frac{\partial|\psi^R_m(t)\rangle}{\partial t}=H(t)|\psi^R_m(t)\rangle,\\
i\frac{\partial|\psi^L_m(t)\rangle}{\partial t}=H^\dagger(t)|\psi^L_m(t)\rangle
\end{align}
\end{subequations}
where $1\leq m\leq N$. Suppose $t_0$ is the starting point of the evolution and $\langle\psi_m^L(t_0)|\psi_n^R(t_0)\rangle=\delta_{mn}$, then the dual evolution preserves the biorthogonal relation $\langle\psi_m^L(t)|\psi_n^R(t)\rangle=\delta_{mn}$, since the two components of its time derivative cancel each other: $\partial_t\langle\psi_m^L|\psi_n^R\rangle=\langle\dot\psi_m^L|\psi_n^R\rangle+\langle\psi_m^L|\dot\psi_n^R\rangle
=i\langle\psi_m^L|H|\psi_n^R\rangle-i\langle\psi_m^L|H|\psi_n^R\rangle=0$. In contrast, the conventional state norms $\langle\psi_m^R|\psi_m^R\rangle$ and $\langle\psi_m^L|\psi_m^L\rangle$ can vary with time. Moreover, the projectors formulated by both left and right states constitute a resolution of the identity operator: $\sum_{n=1}^N|\psi_n^R(t)\rangle\langle\psi_n^L(t)|=\sum_{n=1}^N|\psi_n^L(t)\rangle\langle\psi_n^R(t)|=\mathcal{I}$, that conserve the completed condition. Biorthogonality provides an algebraic structure similar to orthogonality in the Hermitian quantum mechanics by exploiting the dual states in Eq.~(\ref{eq:bioshro}).

From the perspective of UQC, the system dynamics can be described in an ancillary rotating frame spanned by the ket and bra bases $\{|\mu_m^R(t)\rangle, |\mu_m^L(t)\rangle\}$ that also satisfy the biorthogonal and completed conditions, i.e., $\langle\mu^L_m(t)|\mu^R_n(t)\rangle=\delta_{mn}$ and $\sum_{n=1}^N|\mu^R_n(t)\rangle\langle\mu^L_n(t)|=\mathcal{I}$. Then we consider the transformation from the time-dependent biorthogonal basis to the stationary biorthogonal basis $\{|\mu_k^R(t_0)\rangle,|\mu_k^L(t_0)\rangle\}$ with rotation operator $\mathcal{V}(t)\equiv\sum_{k=1}^N|\mu_k^R(t)\rangle\langle\mu_k^L(t_0)|$. Note $\mathcal{V}(t_0)=\mathcal{I}$. This transformation is generally nonunitary and its inverse $\mathcal{V}^{-1}(t)=\sum_{k=1}^N|\mu_k^R(t_0)\rangle\langle\mu_k^L(t)|$, rather than its adjoint, should be applied to the ket states with $|\psi_m^R(t)\rangle_{\rm rot}\equiv\mathcal{V}^{-1}(t)|\psi_m^R(t)\rangle$. The rotated Hamiltonian is thus obtained by a similarity transformation instead of a unitary transformation, i.e., $\mathcal{V}^{\dagger}(t)=\sum_{k=1}^N|\mu_k^L(t_0)\rangle\langle\mu_k^R(t)|$ is replaced by $\mathcal{V}^{-1}(t)$,
\begin{equation}
\begin{aligned}\label{eq:Hrot}
&H_{\rm rot}(t)\equiv\mathcal{V}^{-1}(t)H\mathcal{V}(t)-i\mathcal{V}^{-1}(t)\dot{\mathcal{V}}(t)\\
&=\sum_{k,m=1}^N[H_{km}(t)-\mathcal{A}_{km}(t)]|\mu_k^R(t_0)\rangle\langle\mu_m^L(t_0)|\\
&=\mathcal{V}^{-1}(t)[H(t)-\mathcal{A}(t)]\mathcal{V}(t).
\end{aligned}
\end{equation}
Here $H_{km}(t)\equiv\langle\mu_k^L(t)|H(t)|\mu_m^R(t)\rangle$ and $\mathcal{A}_{km}(t)\equiv i\langle\mu_k^L(t)|\dot\mu_m^R(t)\rangle$ denote the matrix elements of the non-Hermitian Hamiltonian and the non-Hermitian gauge potential $\mathcal{A}(t)$, respectively. Note in the context of the similarity transformation, the Hamiltonian matrix $H_{\rm rot}(t)$ and its conjugate are spanned by $\{\langle\mu_k^L(t_0)|, |\mu_k^R(t_0)\rangle\}$ and $\{\langle\mu_k^R(t_0)|, |\mu_k^L(t_0)\rangle\}$, respectively, where the row basis is not conjugate to the column basis. In addition, the left states are transformed to $|\psi_m^L(t)\rangle_{\rm rot}\equiv\mathcal{V}^\dagger(t)|\psi_m^L(t)\rangle$ in the rotating picture, so that
\begin{equation}
\begin{aligned}
i\partial_t|\psi_m^L\rangle_{\rm rot}&=\left[\mathcal{V}^\dagger H^\dagger(\mathcal{V}^\dagger)^{-1}+i\dot{\mathcal{V}}^\dagger(\mathcal{V}^\dagger)^{-1}\right]|\psi_m^L\rangle_{\rm rot}\\ &=H_{\rm rot}^\dagger|\psi_m^L\rangle_{\rm rot}.
\end{aligned}
\end{equation}
Then the dual Schr\"odinger equation and the biorthogonal relation $\langle\psi_m^L|\psi_n^R\rangle_{\rm rot}=\langle\psi_m^L|\psi_n^R\rangle$ are retained.

The gauge potential can be expressed directly through the ancillary paths:
\begin{equation}\label{eq:gauge-potential}
\mathcal{A}(t)=i\sum_{k,m=1}^N\langle\mu_k^L|\dot\mu_m^R\rangle|\mu_k^R\rangle\langle\mu_m^L|=i\sum_{m=1}^N|\dot\mu_m^R\rangle\langle\mu_m^L|.
\end{equation}
It describes the purely geometric kinematics of the ancillary frame $i\partial_t|\mu_k^R(t)\rangle=\mathcal{A}(t)|\mu_k^R(t)\rangle$, which is not necessarily relevant to the system Hamiltonian $H(t)$. Using $\langle\dot\mu_m^R|\mu_n^L\rangle=-\langle\mu_m^R|\dot\mu_n^L\rangle$, the conjugate of the complex potential $\mathcal{A}(t)$ is found to be
\begin{equation}\label{eq:gauge-potentialconjugate}
\begin{aligned}
&\mathcal{A}^\dagger(t)=\sum_{k=1}^N\sum_{m=1}^N-i\langle\dot{\mu}_m^R(t)|\mu_k^L(t)\rangle|\mu_m^L(t)\rangle\langle\mu_k^R(t)|\\
&=\sum_{k=1}^N\sum_{m=1}^Ni\langle\mu_m^R(t)|\dot{\mu}_k^L(t)\rangle|\mu_m^L(t)\rangle\langle\mu_k^R(t)|\\
&=\sum_{k=1}^Ni|\dot{\mu}^L_k(t)\rangle\langle\mu^R_k(t)|.
\end{aligned}
\end{equation}
Similarly, we have $i\partial_t|\mu_k^L(t)\rangle=\mathcal{A}^\dagger(t)|\mu_k^L(t)\rangle$. The equation of motions for system basis states in the right and left spaces are therefore generated by dual operators, analog to the state evolutions generated by $H$ and $H^\dagger$, respectively. With Eq.~(\ref{eq:Hrot}), the Schr\"odinger equations in the rotating picture are consistent with the original equations~(\ref{eq:bioshro}):
\begin{subequations}
\begin{align}
i\frac{\partial}{\partial t}|\psi_m^R(t)\rangle_{\rm rot}&=H_{\rm rot}(t)|\psi_m^R(t)\rangle_{\rm rot},\\
i\frac{\partial}{\partial t}|\psi_m^L(t)\rangle_{\rm rot}&=H^\dagger_{\rm rot}(t)|\psi_m^L(t)\rangle_{\rm rot}.
\end{align}
\end{subequations}

Assuming that $H_{\rm rot}(t)$ can be partially diagonal in the subspace spanned by $\tilde N$-pairs of selected ancillary biorthogonal basis, $\tilde N\leq N$, then the time-evolution operator spanned by $\{\langle\mu_k^L(t_0)|, |\mu_k^R(t_0)\rangle\}$ can be written as,
\begin{equation}
\tilde{U}^R_{\rm rot}(t)=\sum_{k=1}^{\tilde{N}}e^{-if_k(t)}|\mu^R_k(t_0)\rangle\langle\mu^L_k(t_0)|,
\end{equation}
where $f_k(t)=\int_{t_0}^{t}[H_{kk}(t')-\mathcal{A}_{kk}(t')]\,d t'$ is the accumulated complex phase of the $k$th transitionless pathway. The real part of $f_k$ represents a phase rotation and the imaginary part describes amplification or attenuation of the conventional state norm. Correspondingly, the time-evolution operator spanned by $\{\langle\mu_k^R(t_0)|, |\mu_k^L(t_0)\rangle\}$ is
\begin{equation}
\tilde{U}^L_{\rm rot}(t)=\sum_{k=1}^{\tilde{N}}e^{-if^*_k(t)}|\mu^L_k(t_0)\rangle\langle\mu^R_k(t_0)|.
\end{equation}
$\tilde{N}=N$ means that the whole Hilbert space of the system is under control and the time-dependent non-Hermitian system is completely solvable. In this case, we have $U^R_{\rm rot}(t)$ and $U^L_{\rm rot}(t)$ in full rank. Returning to the original representation, the dual time-evolution operators read
\begin{subequations}\label{eq:dual_evolution_operators}
\begin{align}
U_R(t)=\mathcal{V}(t)U^R_{\rm rot}(t)=\sum_{k=1}^{N}e^{-if_k(t)}|\mu^R_k(t)\rangle\langle\mu^L_k(t_0)|,\\
U_L(t)=[\mathcal{V}^\dagger(t)]^{-1}U^L_{\rm rot}(t)=\sum_{k=1}^{N}e^{-if^*_k(t)}|\mu^L_k(t)\rangle\langle\mu^R_k(t_0)|.
\end{align}
\end{subequations}
Either propagator is not unitary in general since the right and left ancillary states are not necessary orthogonal to each other under the conventional inner product.
Due to the completed condition for biorthogonality, we have $U^\dagger_{L,R}(t)U_{R,L}(t)=\mathcal{I}$, which ensures the conservation of the biorthogonal inner products $\langle\psi_m^L(t)|\psi_n^R(t)\rangle=\delta_{mn}$.

\emph{Main result.} Here, we prove that the von Neumann equation
\begin{equation}\label{von_Neumann}
\frac{\partial}{\partial t}\Pi_k(t)=-i[H(t), \Pi_k(t)], \quad 1\leq k\leq \tilde{N},
\end{equation}
where $\Pi_k(t)\equiv|\mu_k^R(t)\rangle\langle\mu_k^L(t)|$ is the biorthogonal ancillary projector, is the necessary and sufficient condition for the diagonalization of $H_{\rm rot}(t)$ with nonvanishing diagonal elements in the subspace of $\tilde N$ dimensions,  $1\leq\tilde N\leq N$. In this subspace, the vanishing off-diagonal elements of the rotated Hamiltonian $H_{\rm tot}(t)$ reflect the neutralization of the time-dependent system Hamiltonian and the gauge potential induced by frame rotation.

\emph{Necessary condition.} For a rotated Hamiltonian $H_{\rm rot}(t)$ that is diagonal in the ancillary biorthogonal basis $\{\langle\mu_k^L(t)|, |\mu_k^R(t)\rangle\}$, we have $[H_{\rm rot}(t), \Pi_k(t_0)]=0$, i.e.,
\begin{equation}\label{Nece_step1}
\begin{aligned}
&\mathcal{V}^{-1}(t)[H(t)-\mathcal{A}(t)]\mathcal{V}(t)\Pi_k(t_0)\\
&\quad=\Pi_k(t_0)\mathcal{V}^{-1}(t)[H(t)-\mathcal{A}(t)]\mathcal{V}(t).
\end{aligned}
\end{equation}
Multiplying Eq.~(\ref{Nece_step1}) with $\mathcal{V}(t)$ from the left and $\mathcal{V}^{-1}(t)$ from the right, we have
\begin{equation}
\begin{aligned}
&[H(t)-\mathcal{A}(t)]\mathcal{V}(t)\Pi_k(t_0)\mathcal{V}^{-1}(t)\\
&\quad=\mathcal{V}(t)\Pi_k(t_0)\mathcal{V}^{-1}(t)[H(t)-\mathcal{A}(t)],
\end{aligned}
\end{equation}
which is equivalent to $[H(t)-\mathcal{A}(t)]\Pi_k(t)=\Pi_k(t)[H(t)-\mathcal{A}(t)]$ due to the identity $\mathcal{V}(t)\Pi_k(t_0)\mathcal{V}^{-1}(t)=\Pi_k(t)$. More compactly, we have
\begin{equation}\label{Nece_step2}
[\Pi_k(t), \mathcal{A}(t)]=[\Pi_k(t), H(t)].
\end{equation}
Using the dual kinematics associated with gauge potential and its conjugate in Eqs.~(\ref{eq:gauge-potential}) and (\ref{eq:gauge-potentialconjugate}), respectively, we have
\begin{subequations}
\begin{align}
i\frac{\partial}{\partial t}|\mu_k^R(t)\rangle&=\mathcal{A}(t)|\mu_k^R(t)\rangle,\\
-i\frac{\partial}{\partial t}\langle\mu_k^L(t)|&=\langle\mu_k^L(t)|\mathcal{A}(t).
\end{align}
\end{subequations}
Straightforwardly from Eq.~(\ref{Nece_step2}), one can deduce
\begin{equation}
\frac{\partial}{\partial t}\Pi_k(t)=-i[H(t), \Pi_k(t)].
\end{equation}

\emph{Sufficient condition.} The von Neumann equation~(\ref{von_Neumann}) can be rewritten as
\begin{equation}\label{suffi_step1}
i\frac{\partial}{\partial t}[\mathcal{V}(t)\Pi_k(t_0)\mathcal{V}^{-1}(t)]=[H(t),\mathcal{V}(t)\Pi_k(t_0)\mathcal{V}^{-1}(t)].
\end{equation}
Multiplying Eq.~(\ref{suffi_step1}) with $\mathcal{V}^{-1}(t)$ from the left and $\mathcal{V}(t)$ from the right, we have
\begin{equation}
\begin{aligned}
&i\mathcal{V}^{-1}(t)\dot{\mathcal{V}}(t)\Pi_k(t_0)-i\Pi_k(t_0)\mathcal{V}^{-1}(t)\dot{\mathcal{V}}(t)\\
&\quad=\mathcal{V}^{-1}(t)[H(t),\mathcal{V}(t)\Pi_k(t_0)\mathcal{V}^{-1}(t)]\mathcal{V}.
\end{aligned}
\end{equation}
And it turns out to be equivalent to
\begin{equation}
  [H_{\rm rot}, \Pi_k(t_0)]=0,
\end{equation}
by using $\mathcal{V}^{-1}(t)\dot{\mathcal{V}}(t)=-\dot{\mathcal{V}}^{-1}(t)\mathcal{V}(t)$ and Eq.~(\ref{eq:Hrot}). Thus, $H_{\rm rot}(t)$ is diagonal in $\{\langle\mu_k^L(t_0)|, |\mu_k^R(t_0)\rangle\}$.

Briefly for the universal control over non-Hermitian systems, the von Neumann equation~(\ref{von_Neumann}) serves as an exact inverse-engineering prescription. One can first design biorthogonal paths in the rotating frame linking the initial and target states and then for each pathway or basis, solve the rotated Hamiltonian $H_{\rm rot}(t)$ such that $H_{k\neq m}(t)-\mathcal{A}_{k\neq m}(t)=0$, provided that the relevant parameters are available under realistic controls. No adiabatic approximation is involved in the prescription. The remaining diagonal elements of $H_{\rm rot}(t)$ determine the complex phases of the basis. And then extra constraints of the controllable parameters, that the imaginary part of the accumulated phases vanish in the end, can be imposed to yield an effective unitary transformation for the whole passage. These steps underlie the protocols of population transfer and gate construction developed below.

\section{Population transfer in a non-Hermitian two-level system}\label{sec:popu_transfer}

In this section, our theory is applied to the perfect population transfer in an open two-level system described by the dual Schr\"odinger equations~(\ref{eq:bioshro}), where the time-dependent non-Hermitian Hamiltonian reads
\begin{equation}\label{eq:two-level-H}
\begin{aligned}
H(t)&=\frac{\Delta(t)+i\gamma(t)}{2}(|1\rangle\langle 1|-|0\rangle\langle 0|)\\
&\quad+\frac{\Omega(t)}{2}\left[e^{i\phi(t)}|1\rangle\langle 0|+\mathrm{H.c.}\right].
\end{aligned}
\end{equation}
Here the detuning $\Delta(t)$, the level coupling strength or Rabi frequency $\Omega(t)$, the coupling phase $\phi(t)$, and the gain or loss parameter $\gamma(t)$ are real. The ancillary right basis states can be parameterized through an SU(2)-like rotation~\cite{JinJing2025Entangling}:
\begin{subequations}\label{complexalphaR}
\begin{align}\label{muR1}
|\mu^R_1(t)\rangle=\cos\theta(t)e^{i\alpha(t)/2}|0\rangle-\sin\theta(t)e^{-i\alpha(t)/2}|1\rangle, \\
|\mu^R_2(t)\rangle=\sin\theta(t)e^{i\alpha(t)/2}|0\rangle+\cos\theta(t)e^{-i\alpha(t)/2}|1\rangle,
\end{align}
\end{subequations}
where $\theta(t)$ is a real angle and $\alpha(t)=\alpha_r(t)+i\alpha_i(t)$ is a complex phase. $\alpha_i$ vanishes in Hermitian systems, i.e., $\gamma=0$. Correspondingly, the left states are given by
\begin{subequations}\label{complexalphaL}
\begin{align}
|\mu^L_1(t)\rangle=\cos\theta(t)e^{i\alpha^*(t)/2}|0\rangle-\sin\theta(t)e^{-i\alpha^*(t)/2}|1\rangle, \\
|\mu^L_2(t)\rangle=\sin\theta(t)e^{i\alpha^*(t)/2}|0\rangle+\cos\theta(t)e^{-i\alpha^*(t)/2}|1\rangle.
\end{align}
\end{subequations}

Substituting Eqs.~(\ref{complexalphaR}) and (\ref{complexalphaL}) into the von Neumann equation~(\ref{von_Neumann}) for both projectors $\Pi_1(t)$ and $\Pi_2(t)$, that gives rise to the conditions for activating both pairs of useful passages $\{|\mu^R_1(t)\rangle, |\mu^L_1(t)\rangle\}$ and $\{|\mu^R_2(t)\rangle, |\mu^L_2(t)\rangle\}$:
\begin{subequations}
\begin{align}
&\Omega\sin(\alpha_r+\phi)\cosh\alpha_i=-2\dot\theta,\\
&\Omega\cos(\alpha_r+\phi)\sinh\alpha_i=0,\label{Omegacos}\\
&\sin(2\theta)(\dot\alpha_r-\Delta)+\Omega\cos(2\theta)\cos(\alpha_r+\phi)\cosh\alpha_i=0,\\
&\sin(2\theta)(\dot\alpha_i-\gamma)-\Omega\cos(2\theta)\sin(\alpha_r+\phi)\sinh\alpha_i=0.
\end{align}
\end{subequations}
When $\alpha_i\neq0$ and $\Omega\neq0$, Eq.~(\ref{Omegacos}) permits the branch of $\alpha_r+\phi=\pi/2$ (both $\alpha_r$ and $\phi$ are constant in this section). The remaining control conditions are then simplified as
\begin{subequations}
\begin{align}\label{eq:condi_Omega}
\Omega(t)&=-\frac{2\dot\theta(t)}{\cosh\alpha_i(t)},\\
\Delta(t)&=-\dot{\phi}(t),  \label{eq:condi_Delta} \\  \label{eq:condi_gamma}
\gamma(t)&=\dot\alpha_i(t)+2\dot\theta(t)\cot2\theta(t)\tanh\alpha_i(t).
\end{align}
\end{subequations}

Subsequently, the right and left evolution operators can be obtained as
\begin{subequations}
\begin{align}
U_R(t)=\sum_{k=1,2}e^{-if_k(t)}|\mu^R_k(t)\rangle\langle\mu^L_k(t_0)|,\label{TL_EVO_R} \\
U_L(t)=\sum_{k=1,2}e^{-if^*_k(t)}|\mu^L_k(t)\rangle\langle\mu^R_k(t_0)|, \label{TL_EVO_I}
\end{align}
\end{subequations}
where the complex passage phases satisfy
\begin{equation}
\begin{aligned}
&\dot f_{k}(t)=H_{kk}(t)-\mathcal{A}_{kk}(t)=(-1)^{k+1}\times\\
&\left[\frac{\dot\alpha-\Delta-i\gamma}{2}\cos(2\theta)-\frac{\Omega}{2}\sin(2\theta)\cos(\alpha+\phi)\right].
\end{aligned}
\label{eq:transfer-phases}
\end{equation}
Under the conditions in Eqs.~(\ref{eq:condi_Omega}), (\ref{eq:condi_Delta}), and (\ref{eq:condi_gamma}), the time derivatives of the passage phases simplify to
\begin{equation}\label{TL_global_phase}
\dot{f}_{1}(t)=-\frac{i\dot\theta(t)}{\sin2\theta(t)}\tanh\alpha_i(t)=-\dot{f}_{2}(t).
\end{equation}
The time evolutions described by both $U_R(t)$ and $U_L(t)$ are generally nonunitary and $f_k(t)$ has an imaginary part. Nevertheless, $\{|\mu^R_k(t)\rangle, |\mu^L_k(t)\rangle\}$ can be reduced to a set of orthonormal ancillary basis as long as $\alpha_i(T)=0$ at the final time $T$. Together with ${\rm Im}[f_k(T)]=0$, the phase factors in both Eqs.~(\ref{TL_EVO_R}) and (\ref{TL_EVO_I}) are unit in modulus. Consequently, one can have an effective unitary transformation in the end: $U_R^\dagger(T)U_R(T)=U_L^\dagger(T)U_L(T)=\mathcal{I}$. Therefore, an effective coherent control in this non-Hermitian system requires
\begin{align}
&\alpha_i(0)=\alpha_i(T)=0,\\
&\int_0^T\frac{\dot\theta(t)}{\sin[2\theta(t)]}\tanh\alpha_i(t)dt=0. \label{intalphai}
\end{align}
With a given $\theta(t)$, $\alpha_i(t)$ becomes the remaining degree of freedom to cancel the imaginary part of the phase.

To realize a perfect population transfer via the first right passage $|\mu^R_1(t)\rangle$, we choose $\theta(0)=0$ and $\theta(T)=\pi/2$ so that the system can start from $|0\rangle$ to $|1\rangle$ up to a phase. Under the same condition, the second right passage $|\mu^R_2(t)\rangle$ starts from $|1\rangle$ to $|0\rangle$; and $H^\dagger(t)$ drives the two left passages in parallel. For all the ancillary passages simulated in this section, the time-dependent parameters take the same forms as
\begin{align}
\theta(t)&=\frac{\pi}{2}\sin\left(\frac{\pi t}{2T}\right),\label{eq:twolevel_theta}\\
\alpha_i(t)&=A\left[\sin\left(\frac{2\pi t}{T}\right)+b\sin\left(\frac{\pi t}{T}\right)\right], \label{eq:twolevel_alphai}
\end{align}
where the dimensionless coefficient $A$ becomes vanishing in the Hermitian systems and $b$ can be numerically determined by Eq.~(\ref{intalphai}). Here using $T=1$, $\Delta=0$, and $A=0.1$, we have $b\approx0.3434$.

\begin{figure}[htbp]
\centering
\includegraphics[width=0.9\linewidth]{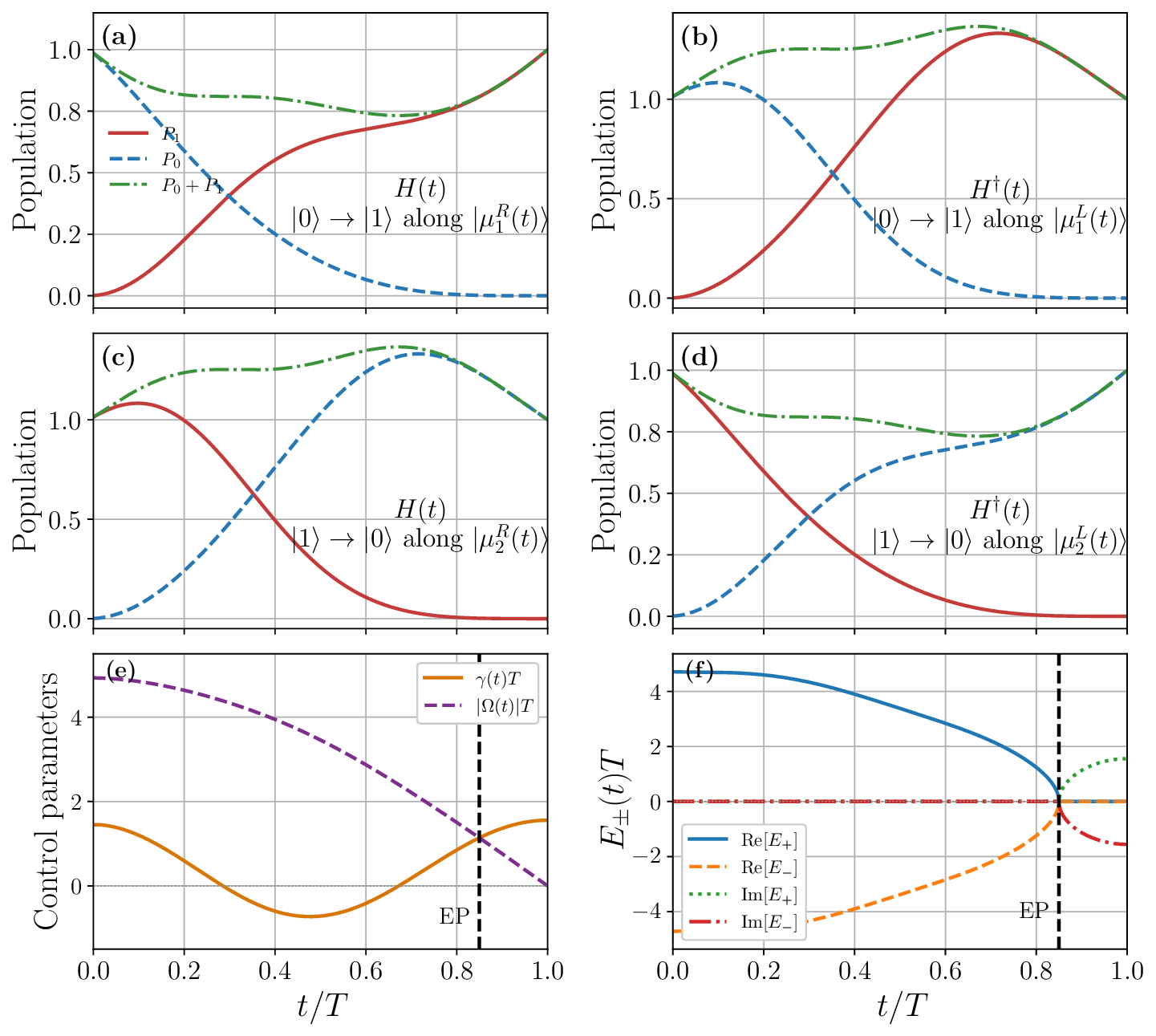}
\caption{Population transfer along both pairs of biorthogonal passages under the same control conditions. (a) and (c): right-state evolution under $H(t)$ from $|0\rangle$ to $|1\rangle$ and from $|1\rangle$ to $|0\rangle$, respectively. (b) and (d): the corresponding left-state evolutions under $H^\dagger(t)$. (e): the time profile of the Rabi frequency $|\Omega(t)|T$ and the gain or loss rate $\gamma(t)T$ determined by Eqs.~(\ref{eq:condi_Omega}) and (\ref{eq:condi_gamma}), respectively. (f): the real and imaginary parts of the instantaneous eigenvalues $E_\pm(t)$ of $H(t)$. The vertical dashed lines in (e) and (f) mark the exceptional point. The bare-level populations $P_0$ and $P_1$ are calculated with unnormalized wavefunctions; and their sum gives the conventional norm. Here $\alpha_r=0$ and $\phi=\pi/2$. $\theta(t)$ and $\alpha_i(t)$ are evaluated by Eqs.~(\ref{eq:twolevel_theta}) and (\ref{eq:twolevel_alphai}), respectively, with $T=1$, $\Delta=0$, $A=0.1$, and $b\approx 0.3434$ that is numerically attained.}
\label{fig:population_transfer}
\end{figure}

\begin{figure}[htbp]
\centering
\includegraphics[width=0.9\linewidth]{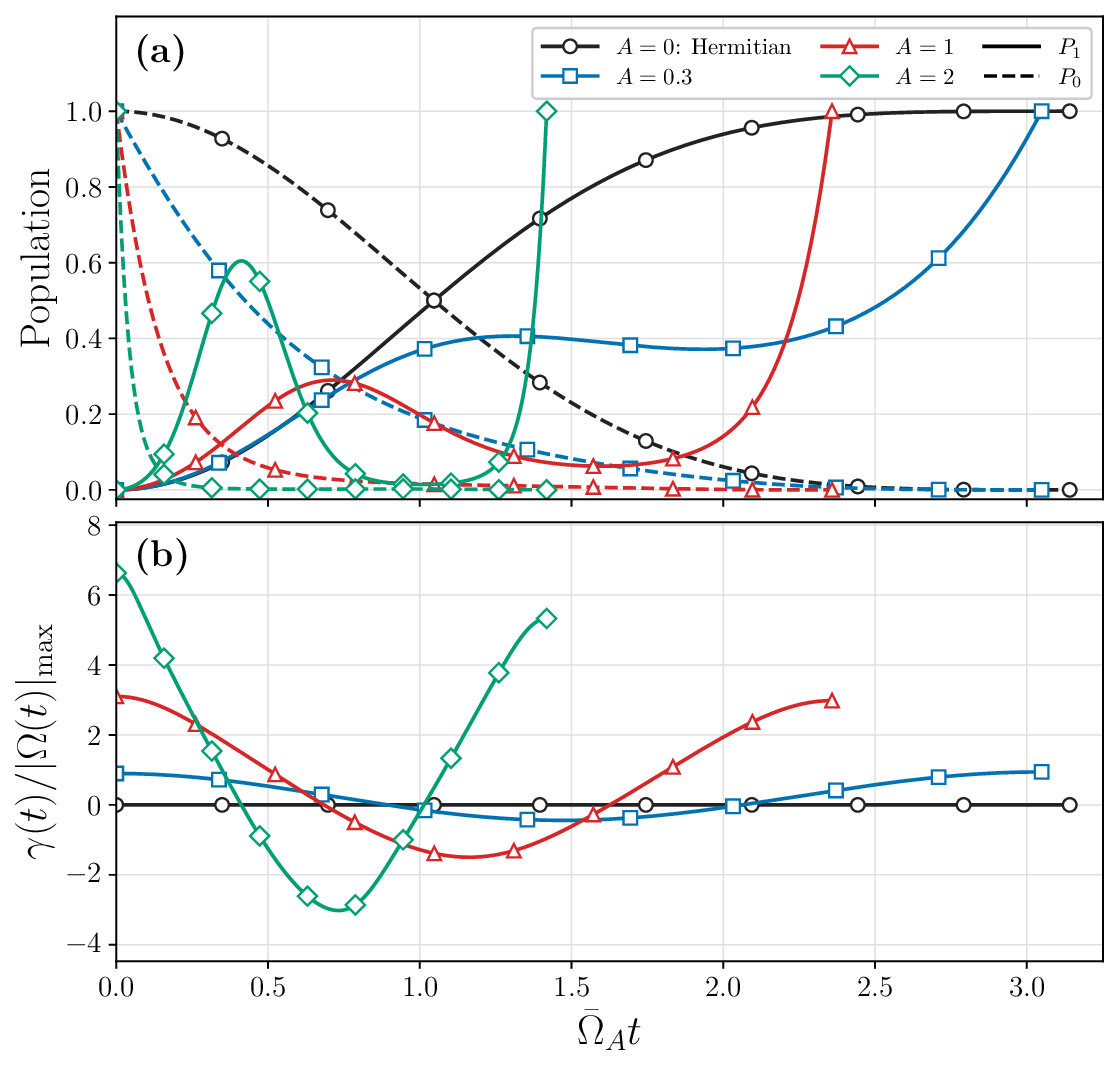}
\caption{Evolution speed of the perfect population transfer under various coefficient $A$'s for the phase $\alpha_i(t)$ in Eq.~(\ref{eq:twolevel_alphai}). (a): Dynamics of populations on the bare levels $|1\rangle$ (dashed lines) and $|0\rangle$ (solid lines) as functions of the time scaled by the averaged Rabi frequency $\bar{\Omega}_A$ (b): Scaled gain or loss rate $\gamma(t)/|\Omega(t)|_{\max}$, where $|\Omega(t)|_{\max}$ is the peak magnitude of Rabi frequency of the driving field. In both (a) and (b), black circles, blue squares, red triangles, and green diamonds mark $A=0$, $0.3$, $1$, and $2$, respectively. Other parameters are the same as in Fig.~\ref{fig:population_transfer}(a). }
\label{fig:speed_advantage}
\end{figure}

In Figs.~\ref{fig:population_transfer}(a), (b), (d), and (d), we plot the bare-level populations $P_j(t)=|\langle j|\psi(t)\rangle|^2$ with $j=0,1$, where $|\psi(t)\rangle=e^{-if_k(t)}|\mu^R_k(t)\rangle$ or $e^{-if^*_k(t)}|\mu^L_k(t)\rangle$ with $k=1,2$. Numerically all the wavefunctions are not artificially normalized during the time evolution. Under exactly the same Hamiltonian $H(t)$, bidirectional population transfers $|0\rangle\rightarrow|1\rangle$ and $|1\rangle\rightarrow|0\rangle$ can be realized along the two right states as shown in Figs.~\ref{fig:population_transfer}(a) and \ref{fig:population_transfer}(c), respectively. The corresponding left-state evolutions under $H^\dagger(t)$ are shown in Figs.~\ref{fig:population_transfer}(b) and \ref{fig:population_transfer}(d). With either initial state, both perfect population transfer and the state-normalization are simultaneously accomplished at $t=T$ in both ket and bra spaces, which indicates that the right and left states coalesce in the end. Also the conditions $\alpha_i(0)=\alpha_i(T)=0$ and ${\rm Im}[f_k(T)]=0$ ensure $U_{R,L}^\dagger(T)U_{R,L}(T)=\mathcal{I}$, so that the norm recovery applies to any initial state as arbitrary superposition of $|0\rangle$ and $|1\rangle$. Figure~\ref{fig:population_transfer}(e) demonstrates the time profiles of $|\Omega(t)|$ and $\gamma(t)$, that support the controlled dynamics and evaluated by Eqs.~(\ref{eq:condi_Omega}) and (\ref{eq:condi_gamma}), respectively. The profile of the gain or loss rate $\gamma(t)$ is subject to the cancellation of the accumulated imaginary part of the global phase of prescribed ancillary passage. The sign of $\gamma(t)$ changes during the passage, reversing the gain or loss assigned to the two bare levels $|0\rangle$ and $|1\rangle$. The Rabi frequency $\Omega(t)$ of the external driving field varies smoothly and vanishes at $t=T$. The vertical dashed line indicates the exceptional point for the time-dependent Hamiltonian~(\ref{eq:two-level-H}), that emerges when $\Delta=0$ and $\Omega^2(t)=\gamma^2(t)\neq0$ and is found to be around $t/T\approx0.85$ under the current setting.

The instantaneous spectrum in Fig.~\ref{fig:population_transfer}(f) confirmes the EP crossing during the population transfer. The eigenvalues of the Hamiltonian~(\ref{eq:two-level-H}) are
\begin{equation}\label{eigenvalues}
  E_{\pm}(t)=\pm\sqrt{D(t)}/2,
\end{equation}
where $D(t)\equiv\Omega^2(t)+[\Delta(t)+i\gamma(t)]^2$. With $\Delta=0$, the eigenvalues in Fig.~\ref{fig:population_transfer}(f) are real when $|\Omega(t)|>|\gamma(t)|$, become purely imaginary when $|\Omega(t)|<|\gamma(t)|$, and coalesce with each other when $|\Omega(t)|=|\gamma(t)|\neq0$. In the presence of EP, the populations in Figs.~\ref{fig:population_transfer}(a)--\ref{fig:population_transfer}(d) evolve smoothly and perfect transfer with norm recovery is retained at $t=T$. Due to the fact that the controlled passages are prescribed by the biorthogonal ancillary bases and the von Neumann equation~(\ref{von_Neumann}), rather than by the adiabatic following of the instantaneous eigenstates of $H(t)$, the coalescence of the eigenvectors does not disturb the controlled dynamics.

Equation~(\ref{eq:condi_Omega}) suggests that for a given $\theta(t)$, a nonzero $\alpha_i(t)$ would reduce the magnitude of the Rabi frequency $\Omega(t)$ by a less-than-unit factor $1/\cosh\alpha_i(t)<1$. In contrast to the Hermitian case with a vanishing $\alpha_i(t)$, a controllable gain or loss rate $\gamma(t)$ in the non-Hermitian Hamiltonian~(\ref{eq:two-level-H}) can be used as an extra resource to accelerate the population-transfer speed with respect to the same averaged Rabi frequency $\bar{\Omega}_A=T^{-1}\int_0^T|\Omega(t)|dt$, where the subscript $A$ implies the dependence on $A$ for $\alpha_i(t)$ in Eq.~(\ref{eq:twolevel_alphai}).

For the Hermitian case, the relative phase $\alpha(t)$ in the ancillary basis, e.g., Eq.~(\ref{muR1}), is real, i.e., $\alpha_i(t)=0$ or $A=0$. The time integral of the control field is $\int_0^T|\Omega(t)|dt=\pi$ for the perfect population transfer from $|0\rangle$ to $|1\rangle$ or from $|1\rangle$ to $|0\rangle$, under the resonant condition $\Delta=0$ and an invariant $\phi$. To quantify the reduction in control area by the factor $1/\cosh\alpha_i(t)$ deduced in Eq.~(\ref{eq:condi_Omega}), which is essentially a reduction in the external driving energy, we evaluate in Fig.~\ref{fig:speed_advantage}(a) the perfect population-transfer process along the first right-state passage $|\mu^R_1(t)\rangle$ against the scaled evolution time $\bar{\Omega}_At$. We compare the transfer protocols under different magnitudes of $\alpha_i(t)$ that are roughly proportional to the coefficient $A$. In particular, we use $A=0$ (the Hermitian case), $0.3$, $1$, and $2$, that yields various $b$'s due to Eqs.~(\ref{intalphai}) and (\ref{eq:twolevel_alphai}). It is found that a larger $A$ results in a quicker population transfer, which confirms that the controlled gain or loss in the non-Hermitian Hamiltonian can accelerate the evolution under the same averaged external driving~\cite{Torosov2014PRA,PRL2007Bender,JPA2012Uzdin}. The profiles of gain or loss rate are shown in Fig.~\ref{fig:speed_advantage}(b) with various $A$'s. A larger $A$ suggests a wider regime of $\gamma(t)$ under control. In other words, the magnitude of gain or loss compensates the coherent driving power. The advantage in speed is therefore a result of the interplay between coherent control and controlled gain or loss, while conserving the unitary transformation between endpoints.

\section{Non-Hermitian holonomic quantum gate}\label{sec:gate}

Under the biorthogonal framework, we can apply UQC to implement a non-Hermitian holonomic quantum gate in a time-dependent non-Hermitian three-level system of a typical $\Lambda$ type. The ground bare-levels $|0\rangle$ and $|1\rangle$ span the computational subspace $\{|0\rangle, |1\rangle\}$ and simultaneously couple to the excited state $|e\rangle$ by external driving. Then the full Hamiltonian reads
\begin{equation}\label{eq:H}
\begin{aligned}
H(t)&=[\Delta(t)-i\gamma(t)]|e\rangle\langle e|+\frac{1}{2}\Big\{\Omega_0(t)e^{-i\beta(t)}|e\rangle\langle 0|
\\&+\Omega_1(t)e^{-i[\beta(t)+\phi(t)]}|e\rangle\langle 1|+\mathrm{H.c.}\Big\},
\end{aligned}
\end{equation}
where $\Delta(t)$ is the excited-state detuning, and $\gamma(t)$ effectively describes loss for $\gamma>0$ and gain for $\gamma<0$. The real amplitudes $\Omega_0(t)$ and $\Omega_1(t)$ are two Rabi frequencies with a common phase $\beta(t)$ and a relative phase $\phi(t)$. The right ancillary bases can be formulated as
\begin{subequations}
\begin{align}
|\mu_1^R(t)\rangle&=\cos\theta|0\rangle+e^{i\phi_0}\sin\theta|1\rangle,\\
|\mu_2^R(t)\rangle&=\cos\chi(t)|b_R(t)\rangle+e^{i\alpha(t)}\sin\chi(t)|e\rangle,\\
|\mu_3^R(t)\rangle&=\sin\chi(t)|b_R(t)\rangle-e^{i\alpha(t)}\cos\chi(t)|e\rangle,
\end{align}
\end{subequations}
where $|b_R(t)\rangle=\sin\theta|0\rangle-e^{i\phi_0}\cos\theta|1\rangle$ is orthogonal to $|\mu_1^R(t)\rangle$. Its dual left bases are then given by
\begin{subequations}
\begin{align}
|\mu_1^L(t)\rangle&=\cos\theta|0\rangle+e^{i\phi_0^*}\sin\theta|1\rangle,\\
|\mu_2^L(t)\rangle&=\cos\chi(t)|b_L(t)\rangle+e^{i\alpha^*(t)}\sin\chi(t)|e\rangle,\\
|\mu_3^L(t)\rangle&=\sin\chi(t)|b_L(t)\rangle-e^{i\alpha^*(t)}\cos\chi(t)|e\rangle,
\end{align}
\end{subequations}
where $|b_L(t)\rangle=\sin\theta|0\rangle-e^{i\phi_0^*}\cos\theta|1\rangle$ is orthogonal to $|\mu_1^L(t)\rangle$. Both $|b_R(t)\rangle$ and $|b_L(t)\rangle$ can be called as bright states.

To construct a holonomic gate in the subspace spanned by $\{|0\rangle, |1\rangle\}$, at least two pairs of passages in both ket and bra spaces should satisfy the von Neumann equation~(\ref{von_Neumann}). On substituting $\Pi_1(t)=|\mu_1^R(t)\rangle\langle\mu_1^L(t)|$ into Eq.~(\ref{von_Neumann}), we can obtain the conditions to activate the nonadiabatic passage pair $|\mu_1^R(t)\rangle$ and $|\mu_1^L(t)\rangle$:
\begin{subequations}
\begin{align}
\dot{\theta}(t)&=\dot{\phi}_0(t)=0,\\
\frac{\Omega_0(t)}{\Omega_1(t)}&=-e^{i\phi_0-i\phi(t)}\tan\theta.
\end{align}
\end{subequations}
It is found that $\theta$ and $\phi_0$ are no longer dependent on time, i.e., they are real constants. We can set $\Omega_0(t)=\Omega(t)\sin\theta$, $\Omega_1(t)=-\Omega(t)\cos\theta$, and $\phi(t)=\phi_0$, consistent with the conventional nonadiabatic holonomic gate protocols~\cite{NJP2012Erik,PRL2019Liu}. Consequently, a pair of static dark states appear as $|\mu_1^{R,L}(t)\rangle=|\mu_1^{R,L}\rangle$, remaining separated from the other states as well as the system dynamics. Subsequently, we substitute $\Pi_2(t)=|\mu_2^R(t)\rangle\langle\mu_2^L(t)|$ to Eq.~(\ref{von_Neumann}) to obtain the rest control conditions
\begin{subequations}
\begin{align}
\beta(t)&=\frac{\pi}{2}-\alpha_r(t),\\ \label{eq:gate-Omega}
\Omega(t)&=-\frac{2\dot\chi(t)}{\cosh\alpha_i(t)},\\
\Delta(t)&=-\dot\alpha_r(t),\\
\gamma(t)&=\dot\alpha_i(t)+2\dot\chi(t)\cot[2\chi(t)]\tanh\alpha_i(t).\label{eq:gate-gamma}
\end{align}
\end{subequations}

In the construction of holonomic gates, the system is driven from the computational subspace to the excited state and then back to it through $|\mu_2^{R,L}(t)\rangle$ by setting
\begin{equation}\label{eq:cycle}
\chi(0)=0,\quad \chi(T/2)=\pi/2, \quad\chi(T)=0.
\end{equation}
Thereby $|\mu_2^{R,L}(T)\rangle=|b_{R,L}(T)\rangle$, on which a holonomic phase is finally encoded. In particular, a complex phase $f_2(t)$ [$f_2^*(t)$] can be generated on the bright state $|b_R(t)\rangle$ [$|b_L(t)\rangle$] relative to the dark state $|\mu_1^R\rangle$ ($|\mu_1^L\rangle$) as indicated in Eq.~(\ref{eq:dual_evolution_operators}) during the whole passage. Using Eq.~(\ref{eq:Hrot}), the global phase can be divided into $f_2(t)=\gamma_d(t)+\gamma_g(t)$, where $\gamma_d(t)=\int_0^TH_{22}(t')dt'$ and $\gamma_g(t)=\int_{0}^t\mathcal{A}_{22}(t')dt'$ with $\mathcal{A}_{22}(t)=-\dot{\alpha}(t)\sin^2\chi(t)$ are the dynamical and geometric components, respectively. It becomes purely geometric when $\gamma_d(T)=0$. In compliance with Eq.~(\ref{eq:cycle}), we therefore impose the cancellation of the integrated dynamical contribution over a parametric cycle by alternatively setting
\begin{align}
\chi(t)&=\frac{\pi}{4}\left[1-\cos\left(\frac{2\pi t}{T}\right)\right],\label{eq:chi}\\
\alpha_i(t)&=-A\sin[2\chi(t)].\label{eq:alphai_gate}
\end{align}
Here, $\alpha_i(t)$ is chosen to avoid singularity in Eq.~(\ref{eq:gate-gamma}). $\chi(t)$ and $\alpha_i(t)$ are even functions under $t\mapsto T-t$, while their time-derivatives are odd. To introduce a desired phase difference $\eta$ between the outgoing stage $[0, T/2]$ and the returning stage $[T/2, T]$, we set $\alpha_r(t)=\eta\Theta(t-T/2)$, where $\Theta(t)$ is the Heaviside step function. Then on each stage,
\begin{equation}\label{eq:H22}
H_{22}(t)=i\left[-\dot\alpha_i(t)\sin^2\chi(t)+\dot\chi(t)\tan\chi(t)\tanh\alpha_i(t)\right]
\end{equation}
according to the similarity transformation in Eq.~(\ref{eq:Hrot}). Upon substituting Eqs.~(\ref{eq:chi}) and (\ref{eq:alphai_gate}) into Eq.~(\ref{eq:H22}), it is confirmed that $H_{22}(T-t)=-H_{22}(t)$ as well as $\int_0^T H_{22}(t)dt=0$. Consequently, the dynamical contribution to the phase of the bright states vanishes after a complete process, although $H_{22}(t)$ is generally nonzero. The complex phase along the passage $|\mu_2^R(t)\rangle$ [$|\mu_2^L(t)\rangle$] is therefore purely geometric and determined by the gauge potential.

\begin{figure}[htbp]
\centering
\includegraphics[width=0.9\linewidth]{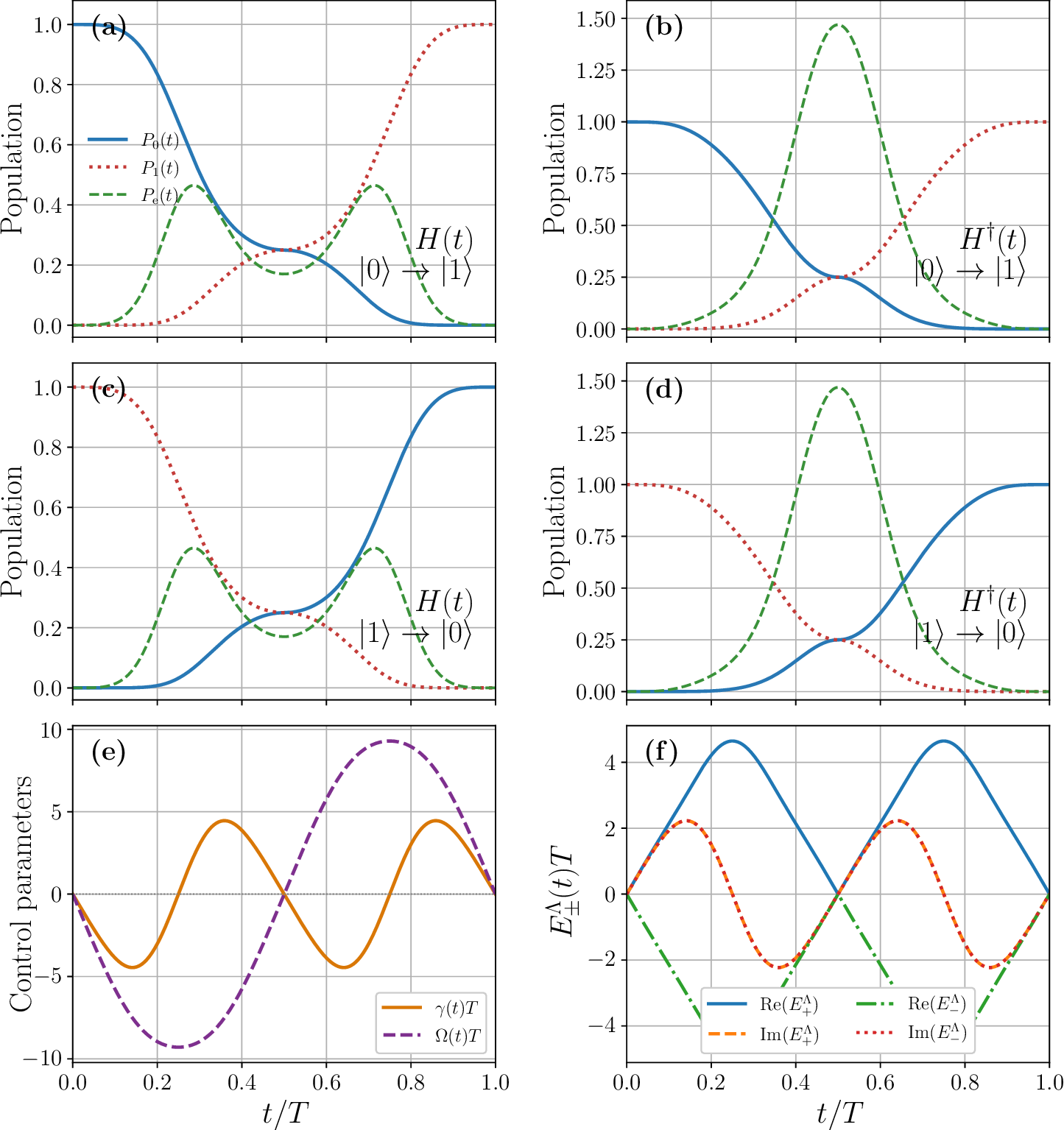}
\caption{The non-Hermitian holonomic $X$-gate in computational space spanned by $\{|0\rangle,|1\rangle\}$. (a) and (c): the population dynamics of the right states under $H(t)$ for system initially on $|0\rangle$ and $|1\rangle$, respectively. (b) and (d): the corresponding left states evolution under $H^\dagger(t)$. (e): the time profiles of the Rabi frequency and gain or loss rate. (f): the real and imaginary parts of the instantaneous eigenvalues. The bare-level populations are calculated by unnormalized wavefunctions. $\chi(t)$ and $\alpha_i(t)$ are determined by Eqs.~(\ref{eq:chi}) and (\ref{eq:alphai_gate}), respectively. $T=1$, $A=0.35$, $\theta=\pi/4$, $\phi_0=0$, and $\eta=\pi$.}
\label{fig:gate}
\end{figure}

To realize a unitary holonomic gate in this open quantum system, the geometric phase $\gamma_g(T)$ should be real at the endpoint. According to Eqs.~(\ref{eq:Hrot}), (\ref{eq:chi}), and (\ref{eq:alphai_gate}), it is found that similar to the dynamical phase, the imaginary part of the gauge term $\mathcal{A}_{22}(t)$ is also symmetrical to $t=T/2$, i.e., ${\rm Im}[\mathcal{A}_{22}(T-t)]=-{\rm Im}[\mathcal{A}_{22}(t)]$. Thus, the imaginary part of the global phase will vanish after a complete running $\int_0^T{\rm Im}[\mathcal{A}_{22}(t)]dt=0$. On the whole, the real geometric phase for this effective unitary holonomic gate can be obtained by
\begin{equation}
\gamma_g=-\int_0^T\dot\alpha_r(t)\sin^2\chi(t)dt=-\eta\sin^2\chi(T/2)=-\eta.
\end{equation}
And the resulting holonomic transformation is
\begin{equation}
\begin{aligned}
U(\theta,\phi_0,\eta)&=|\mu_1^R\rangle\langle\mu_1^L|+e^{-i\eta}|b_R(T)\rangle\langle b_L(T)|\\
&=e^{-i(\eta/2)}e^{i(\eta/2)\hat{n}\cdot\vec{\sigma}},
\end{aligned}
\end{equation}
where $\hat{n}\equiv(\sin 2\theta\cos\phi_0,-\sin2\theta\sin\phi_0,\cos2\theta)$. For $\theta=\pi/4$, $\phi_0=0$, and $\eta=\pi$, the dark and bright components acquire opposite phases, yielding a Pauli $X$ gate.

In Figs.~\ref{fig:gate}(a), (b), (c), and (d), we demonstrate the action of the holonomic $X$-gate on both initial states $|0\rangle$ and $|1\rangle$ in the computational subspace, by plotting the population dynamics on $|i\rangle$ with $i=0,1,e$. It is found that the excited state $|e\rangle$ is temporarily occupied. The dynamics under $H$ and $H^\dagger$ in ket and bra spaces, respectively, are distinct during the gate construction yet ended with the same renormalized point. At $t=T$, the excited-level population completely vanishes and all four passages return to unit in standard norm despite the transient amplification or attenuation. A full-rank gate operation within the computational subspace is therefore established, valid for arbitrary coherent superpositions. Figure~\ref{fig:gate}(e) shows the time profiles of the Rabi frequency and the gain or loss rate obtained by Eqs.~(\ref{eq:gate-Omega}) and (\ref{eq:gate-gamma}), respectively. $\gamma(t)$ alternates in sign, caused by the alternative loss and gain on the excited state, by which the accumulated imaginary part of the passage phase vanishes in the end.

Figure~\ref{fig:gate}(f) shows the corresponding instantaneous spectrum in the two-dimensional subspace spanned by the bright and excited states. On each stage with a constant $\alpha_r$ and $\Delta=0$, the two eigenvalues read $E^\Lambda_\pm(t)=[-i\gamma(t)\pm\sqrt{\Omega^2(t)-\gamma^2(t)}]/2$. With the chosen parameters, $|\Omega(t)|$ always larger than $|\gamma(t)|$, so that $E^\Lambda_\pm(t)$ have opposite real parts and the same imaginary part. Both $\Omega(t)$ and $\gamma(t)$ vanish at the boundary between two stages $t=T/2$, which is also indicated by Fig.~\ref{fig:gate}(e). The resulting degeneracy of the resonant block is not an EP because the two eigenstates do not coalesce. Together with Fig.~\ref{fig:population_transfer}, no significant effect is observed about the EP crossing or avoiding during the prescribed population dynamics or the effective unitary transformation. Our control is fully determined by the ancillary passages and the accumulated phases.

\section{Chiral state conversion}\label{sec:chiral}

Time-dependent parameters tracking a closed path enclosing an exceptional point is known to generate chiral mode conversion, as a characteristic phenomenon in non-Hermitian systems~\cite{Doppler2016Nature,Liu2021PRL}. In particular, the instantaneous eigenstate can exhibit a chiral conversion during time evolution, where the clockwise (CW) and counterclockwise (CCW) direction favor different output state, nearly irrespective to the initial preparation. It is induced by unequal chiral-dependent amplification or attenuation. On the other hand, chiral state conversion typically emerges around EP, where eigenvalues and their associated eigenstates coalesce in the same time~\cite{Doppler2016Nature,Liu2021PRL}. Yet recently, it is found that encircling the exceptional point is not necessarily required~\cite{PRA2017Hassan, Nasari2022Nasari}.

This phenomenon can be conveniently described within our framework of controlled nonunitary dynamics when we hold the imaginary parts of the accumulated phases of nonadiabatic passages $|\mu_{k}^{R,L}(t)\rangle$ instead of imposing an effective unitary transformation between endpoints. The resulting amplification or attenuation of certain biorthogonal passages provides a natural route to the direction dependence of the output state. We use the two-level-system Hamiltonian~(\ref{eq:two-level-H}) with a constant $\gamma(t)=\gamma_0>0$ and $\phi(t)=0$, whose eigenvalues are given by Eq.~(\ref{eigenvalues}) and the instantaneous right and left eigenstates read,
\begin{subequations}\label{eq:chiral_eigenstates}
\begin{align}
|\varphi_-^R(t)\rangle&=\cos\vartheta(t)|0\rangle-\sin\vartheta(t)|1\rangle,\\
|\varphi_+^R(t)\rangle&=\sin\vartheta(t)|0\rangle+\cos\vartheta(t)|1\rangle,\\
|\varphi_-^L(t)\rangle&=\cos\vartheta^*(t)|0\rangle-\sin\vartheta^*(t)|1\rangle,\\
|\varphi_+^L(t)\rangle&=\sin\vartheta^*(t)|0\rangle+\cos\vartheta^*(t)|1\rangle,
\end{align}
\end{subequations}
where the complex mixing angle $\vartheta(t)=\vartheta_r(t)+i\vartheta_i(t)$ is defined by
\begin{equation}
\cos[2\vartheta(t)]=\frac{\Delta(t)+i\gamma_0}{\sqrt{D(t)}},\quad
\sin[2\vartheta(t)]=\frac{\Omega(t)}{\sqrt{D(t)}},
\end{equation}
These eigenstates satisfy $H(t)|\varphi_\pm^R(t)\rangle=E_\pm(t)|\varphi_\pm^R(t)\rangle$ and $H^\dagger(t)|\varphi_\pm^L(t)\rangle=E_\pm^*(t)|\varphi_\pm^L(t)\rangle$ with $\langle\varphi_j^L(t)|\varphi_k^R(t)\rangle=\delta_{jk}$ with $j,k\in\{+,-\}$, while their ordinary norms are not unit in general.

To construct passages that become the instantaneous biorthogonal eigenstates parameterized in Eq.~(\ref{eq:chiral_eigenstates}) at both $t=0$ and $t=T$, we have to use the ancillary biorthogonal ancillary bases with a complex angle $\theta(t)=\theta_r(t)+i\theta_i(t)$ and a real phase $\alpha(t)$~\cite{Ibanez2011} instead of Eq.~(\ref{complexalphaR}). The right basis states are then given by
\begin{subequations}\label{muRcomplextheta}
\begin{align}
|\mu^R_1(t)\rangle&=\cos\theta(t)e^{i\alpha(t)/2}|0\rangle-\sin\theta(t)e^{-i\alpha(t)/2}|1\rangle, \\
|\mu^R_2(t)\rangle&=\sin\theta(t)e^{i\alpha(t)/2}|0\rangle+\cos\theta(t)e^{-i\alpha(t)/2}|1\rangle,
\end{align}
\end{subequations}
and the corresponding left basis states read
\begin{subequations}\label{muLcomplextheta}
\begin{align}
|\mu^L_1(t)\rangle&=\cos\theta^*(t)e^{i\alpha(t)/2}|0\rangle-\sin\theta^*(t)e^{-i\alpha(t)/2}|1\rangle, \\
|\mu^L_2(t)\rangle&=\sin\theta^*(t)e^{i\alpha(t)/2}|0\rangle+\cos\theta^*(t)e^{-i\alpha(t)/2}|1\rangle.
\end{align}
\end{subequations}
Substituting them into the von Neumann equation~(\ref{von_Neumann}), we obtain the constraint conditions:
\begin{subequations}
\begin{align}\label{eq:chiral_control_conditionsOmega}
\Omega(t)\sin\alpha(t)&=-2\dot\theta(t),\\ \label{eq:chiral_control_conditionstheta}
[\Delta(t)-\dot{\alpha}(t)+i\gamma_0]\sin[2\theta(t)]&=\Omega(t)\cos\alpha(t)\cos[2\theta(t)].
\end{align}
\end{subequations}
The left hand side of Eq.~(\ref{eq:chiral_control_conditionsOmega}) is real, which means $\dot\theta_i=0$. We then choose a positive constant $\theta_i$, for which $\sin(2\theta)$ never equals to zero. Separating the real and imaginary parts of Eq.~(\ref{eq:chiral_control_conditionstheta}) gives rise to $\Omega\cos\alpha=-\Omega_0$ and $\Delta-\dot\alpha=\Delta_0$, where
\begin{subequations}\label{eq:chiral_Omega0}
\begin{align}
\Omega_0(t)&\equiv\gamma_0\frac{\cosh(4\theta_i)-\cos[4\theta_r(t)]}{\sinh(4\theta_i)},\\  \label{eq:chiral_Delta0}
\Delta_0(t)&\equiv-\gamma_0\frac{\sin[4\theta_r(t)]}{\sinh(4\theta_i)}.
\end{align}
\end{subequations}
Consequently, the control conditions can be organized as
\begin{subequations}\label{eq:chiral_physical_controls}
\begin{align}
\alpha(t)&=\pi+\arctan\left[\frac{2\dot\theta_r(t)}{\Omega_0(t)}\right],\label{eq:chiral_alpha}\\
\Omega(t)&=\sqrt{\Omega_0^2(t)+4\dot\theta_r^2(t)},\\
\Delta(t)&=\Delta_0(t)+\dot\alpha(t).
\end{align}
\end{subequations}
And the two complex passage phases satisfy
\begin{equation}\label{eq:chiral_passage_phases}
\dot{f}_1(t)=\frac{\gamma_0\sin[2\theta_r(t)]}{2\sinh(2\theta_i)}-i\frac{\gamma_0\cos[2\theta_r(t)]}{2\cosh2\theta_i}=-\dot{f}_2(t).
\end{equation}

Note the instantaneous Hamiltonian can be obtained by inverse transformation over Eq.~(\ref{eq:Hrot}), i.e.,
\begin{equation}\label{eq:chiral_physical_decomposition}
\begin{aligned}
H(t)&=\mathcal V(t)H_{\rm rot}(t)\mathcal V^{-1}(t)
+i\dot{\mathcal V}(t)\mathcal V^{-1}(t)\\
&=\sum_{k=1}^2\dot f_k(t)|\mu_k^R(t)\rangle\langle\mu_k^L(t)|
+\mathcal{A}(t).
\end{aligned}
\end{equation}
Using Eqs.~(\ref{eq:gauge-potential}), (\ref{muRcomplextheta}), and (\ref{muLcomplextheta}) with a positive $\theta_i$ and a real $\alpha$, the off-diagonal gauge terms are found to be 
\begin{equation}\label{eq:chiral_offdiagonal_gauge}
\begin{aligned}
\mathcal{A}_{12}(t)&=-\frac{\dot{\alpha}(t)}{2}\sin[2\theta(t)]+i\dot{\theta}_r(t),\\
\mathcal{A}_{21}(t)&=-\frac{\dot{\alpha}(t)}{2}\sin[2\theta(t)]-i\dot\theta_r(t).
\end{aligned}
\end{equation}
Therefore, the ancillary basis states can become the instantaneous eigenstates of the non-Hermitian system, i.e., $H(T)|\mu_k^R(t)\rangle=\dot{f}_k(T)|\mu_k^R(T)\rangle$ and $H^\dagger(T)|\mu_k^L(T)\rangle=\dot{f}_k^*(T)|\mu_k^L(T)\rangle$, by imposing the boundary constraints $\dot\alpha(0)=\dot\alpha(T)=\dot\theta_r(0)=\dot\theta_r(T)=0$ since $\sin2\theta\neq0$.

In addition, Eqs.~(\ref{eq:chiral_Omega0})--(\ref{eq:chiral_physical_controls}) show that the control parameters $\Delta(t)$ and $\Omega(t)$ are periodic in $\theta_r$ with a period $\pi/2$. We therefore require $\theta_r(T)-\theta_r(0)=\pm\pi/2$ to complete a closed loop in parametric space. In regard of the boundary conditions, we can set 
\begin{equation}\label{eq:chiral_schedule}
\theta_{r,\pm}(t)=-\frac{\pi}{4}\pm\frac{\pi}{2}p(t/T),\quad p(s)\equiv s-\frac{\sin(2\pi s)}{2\pi},
\end{equation}
under which the trajectory in the parametric space is closed with
\begin{equation}\label{eq:chiral_endpoint_controls}
\Omega(0)=\Omega(T)=\gamma_0\coth2\theta_i,\qquad \Delta(0)=\Delta(T)=0.
\end{equation}
Thus, the dual passages coincide with the instantaneous eigenstates of the endpoint Hamiltonian, although they are generally distinct during the time evolution. The imaginary part of the complex passage phases changes sign with the traversal direction:
\begin{equation}\label{eq:chiral_accumulated_gain}
\begin{aligned}
f_{1,\pm}(T)&=\mp i\mathcal{G}=-f_{2,\pm}(T),\\
\mathcal{G}&=\frac{\gamma_0T}{2\cosh2\theta_i}\int_0^1\sin[\pi p(s)]ds>0.
\end{aligned}
\end{equation}
With $\theta(0)=-\pi/4+i\theta_i$, the mutual exchange between passages in the end follows directly from $\theta(T)=\theta(0)\pm\pi/2$:
\begin{equation}\label{eq:chiral_endpoint_exchange}
|\mu_1^R(T)\rangle=\pm|\mu_2^R(0)\rangle, \quad |\mu_2^R(T)\rangle=\mp|\mu_1^R(0)\rangle.
\end{equation}
Then the time-evolution operators in Eq.~(\ref{eq:dual_evolution_operators}) at $t=T$ are modified to be
\begin{equation}\label{eq:chiral_evolution_operators}
\begin{aligned}
U_{R,\pm}(T)=&\pm e^{-if_{1,\pm}(T)}|\mu_2^R(0)\rangle\langle\mu_1^L(0)|\\
&\mp\,e^{-if_{2,\pm}(T)}|\mu_1^R(0)\rangle\langle\mu_2^L(0)|.
\end{aligned}
\end{equation}
Given $|\mu_1^{R,L}(0)\rangle=i|\varphi_-^{R,L}(0)\rangle$ and $|\mu_2^{R,L}(0)\rangle=-i|\varphi_+^{R,L}(0)\rangle$ and Eq.~(\ref{eq:chiral_accumulated_gain}), Eq.~(\ref{eq:chiral_evolution_operators}) can be rewritten as
\begin{equation}\label{eq:chiral_endpoint_maps}
U_{R,\pm}(T)=-e^{\mathcal G}|\varphi_{\mp}^R(0)\rangle\langle\varphi_{\pm}^L(0)|+e^{-\mathcal G}|\varphi_{\pm}^R(0)\rangle\langle\varphi_{\mp}^L(0)|,
\end{equation}
where the subscripts $+$ and $-$ of $U_{R,\pm}(T)$ indicate the chiral state conversion along the counterclockwise and clockwise directions, respectively. Note the amplification factor $\mathcal{G}$ increases with $T$. In the long-time limit, i.e., $T\to\infty$, $|\varphi_{\mp}^R(0)\rangle$ will evolve to $|\varphi_{\pm}^R(0)\rangle$ enforced by the dominated amplified channel. The output state thus depends on the traversal direction.

\begin{figure}[htbp]
\centering
\includegraphics[width=0.9\linewidth]{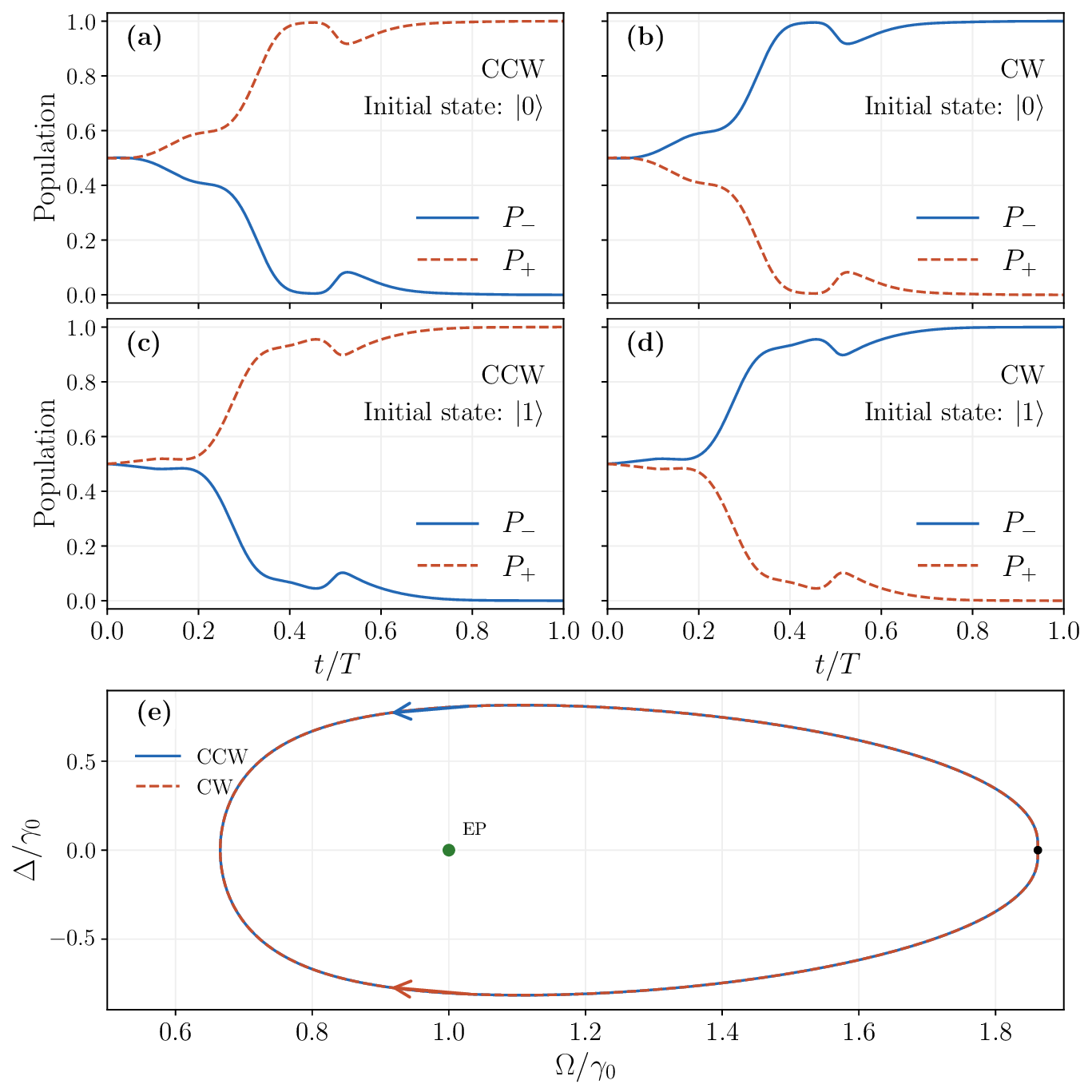}
\caption{Chiral state conversion with EP encircling for $\theta_i=0.3<\theta_{i,c}$. (a) and (b): instantaneous eigenstate populations $P_-(t)$ and $P_+(t)$ for the initial state $|0\rangle$ along the counterclockwise and clockwise loops, respectively. (c) and (d): the relevant results for the initial state $|1\rangle$. (e): the control loops in the space of $(\Omega/\gamma_0, \Delta/\gamma_0)$, as obtained by Eq.~(\ref{eq:chiral_physical_controls}). The green point marks the EP at $(1,0)$ and the black point marks the common starting and ending point. The arrows indicate the two traversal directions. The other parameters are $\theta_r(0)=\pi/4$ and $\gamma_0T=16$. }
\label{fig:chiral_encircling}
\end{figure}

According to the common starting and ending point in Eq.~(\ref{eq:chiral_endpoint_controls}), whether or not the exceptional point $(\Delta/\gamma_0,\Omega/\gamma_0)=(0,1)$ is enclosed by the loops in the parametric space is determined by the location of the loop midpoint that can be controlled by $\theta_i$. By Eqs.~(\ref{eq:chiral_Omega0})--(\ref{eq:chiral_physical_controls}), the midpoint is found to be located at $\Delta(T/2)=0$ and $\Omega(T/2)=\sqrt{\gamma_0^2\tanh^2(2\theta_i)+4\pi^2/T^2}$. Thus, the critical value of $\theta_i$ is determined by $\Omega(T/2)=\gamma_0$, yielding $\theta_{i,c}={\rm arcosh}[\gamma_0T/(2\pi)]/2$.

In Fig.~\ref{fig:chiral_encircling}(a), (b), (c), (d), we demonstrate the chiral state conversion starting from different initial state and along different directions when the loops enclose the EP. Here we use $\gamma_0T=16$, and then it is found that $\theta_{i,c}\approx 0.7934$. The chiral state transfer is described by the relative population on the two nonorthogonal right eigenstates $|\varphi_\pm^R(t)\rangle$, that are defined  as
\begin{equation}\label{eq:chiral_populations}
P_{\pm}(t)=\frac{|\langle\varphi_\pm^L(t)|\psi^R(t)\rangle|^2}{|\langle\varphi_+^L(t)|\psi^R(t)\rangle|^2+|\langle\varphi_-^L(t)|\psi^R(t)\rangle|^2}.
\end{equation}
The two closed control loops in Fig.~\ref{fig:chiral_encircling}(e) enclose the EP without passing through it, so that the instantaneous spectrum remains nondegenerate throughout the evolution. Both bare-level inputs initially have equal relative populations, $P_-(0)=P_+(0)=1/2$ due to $\theta_r(0)=\pi/4$. Under the CCW direction, $|\varphi_+^R(T)\rangle$ dominates after the cycle, as shown in Figs.~\ref{fig:chiral_encircling}(a) and \ref{fig:chiral_encircling}(c). Reversing the direction will amplify the occupation on $|\varphi_-^R(T)\rangle$ instead, as shown in Figs.~\ref{fig:chiral_encircling}(b) and \ref{fig:chiral_encircling}(d). The two initial states renders different dynamics but the same output state for a given direction. In particular, $P_+(T)\approx0.999972$ for CCW direction and $P_-(T)\approx0.999972$ for CW direction. The relative population on the final state is close to unit, although the state propagates without normalization.

\begin{figure}[htbp]
\centering
\includegraphics[width=0.9\linewidth]{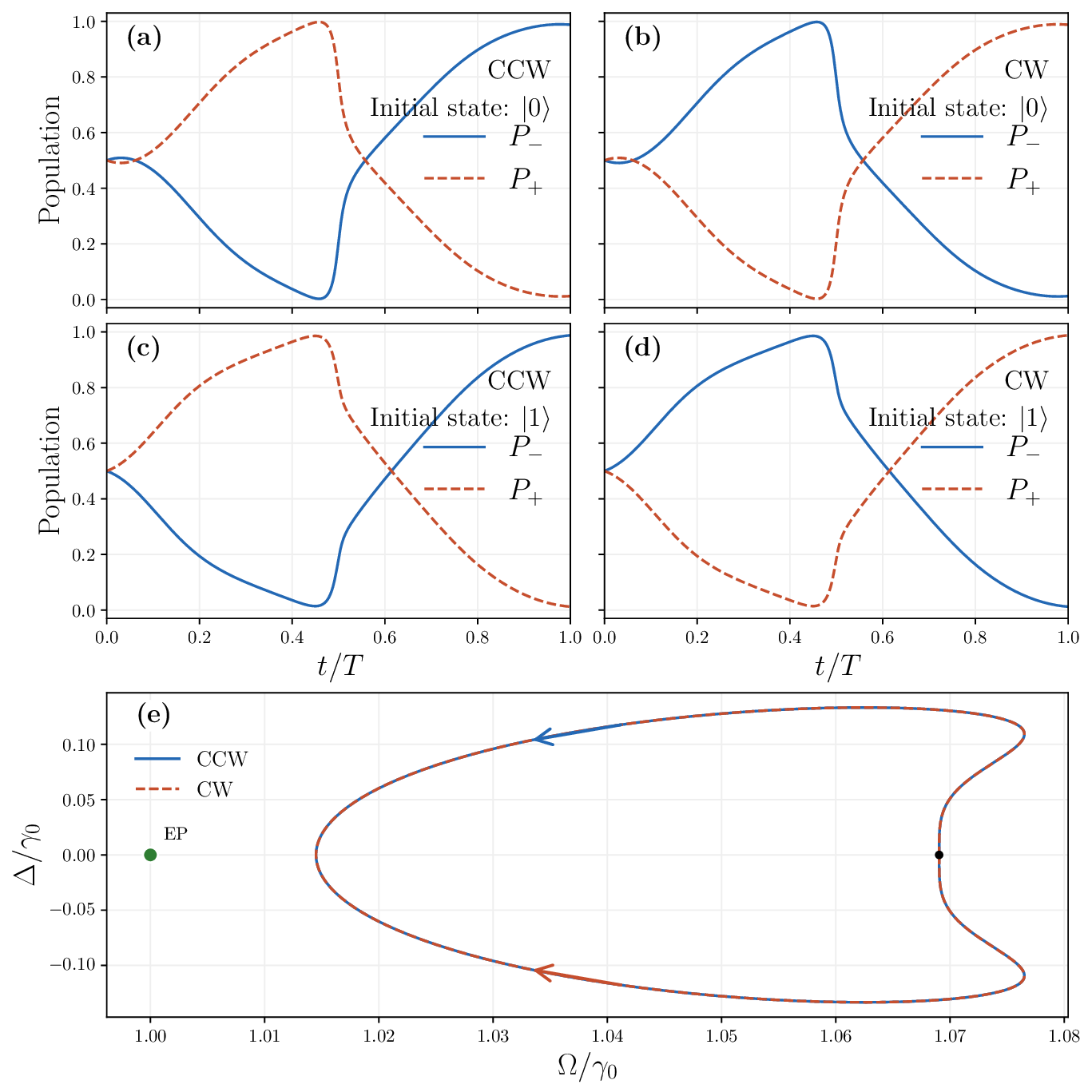}
\caption{Chiral state conversion for $\theta_i=0.85>\theta_{i,c}$. (a) and (b): instantaneous eigenstate populations for the initial state $|0\rangle$ along CCW and CW loops, respectively. (c) and (d): the corresponding results for $|1\rangle$. (e): the control loops that do not enclose the EP. The green and black points mark the EP and the common starting and ending point, respectively. The other parameters are the same as in Fig.~\ref{fig:chiral_encircling}. }
\label{fig:chiral_nonencircling}
\end{figure}

In Fig.~\ref{fig:chiral_nonencircling}(e) for $\theta_i=0.85$, the exceptional point is outside the loops. Nevertheless, in Fig.~\ref{fig:chiral_nonencircling}(a), (b), (c), (d), the output state remains direction-dependent: $|\varphi_-^R(T)\rangle$ dominates during the loop of CCW direction, while $|\varphi_+^R(T)\rangle$ dominates the CW loop, irrespective to the initial state. It suggests that encircling an EP is not necessary for observing the chiral state conversion. 

In comparing Figs.~\ref{fig:chiral_encircling} and \ref{fig:chiral_nonencircling}, the different output state along the same direction can be understood via the Riemann double-sheet structure of the instantaneous spectrum~\cite{Doppler2016Nature,Erdamar2026PRXQ}. A branch cut of the second order is a line along which the two Riemann sheets are joined, enabling a continuously varying eigenvalue as in Eq.~(\ref{eigenvalues}) to pass from one sheet to another. The encircling loop crossing the cut once results in $\sqrt{D(T)}=-\sqrt{D(0)}$ as well as $E_{\pm}(T)\to E_{\mp}(T)$ and the dual eigenstates exchange, i.e., $|\varphi_\pm^R(T)\rangle\to|\varphi_\mp^R(0)\rangle$, which are used to calculate the populations up to an irrelevant phase. The non-encircling loop avoids the cut, holding $\sqrt{D(T)}=\sqrt{D(0)}$. In addition, we find that encircling the EP twice pushes the eigenstates to their original Riemann sheet and then yields the same final state as in the non-encircling protocol.

\section{Discussion and conclusion}\label{sec:conclusion}

The preceding perfect population transfer in the two-level system and the geometric gate in the three-level $\Lambda$ system demand time-dependent control of the gain and loss rates, including changing their signs during the controlled evolution. Optical and magnonic platforms provide feasible routes for the analog simulations over the non-Hermitian systems and dynamics considered in this work at the level of classical mode amplitudes. For example, in the coupled optical fiber loops, Feis et al.~\cite{NatPhoton2025Feis} implemented a discrete-time modulation that alternates amplification and attenuation using erbium-doped fibre amplifiers and acousto-optic modulators. Wu et al.~\cite{NC2026Wu} further demonstrated programmable gain or loss factors in a coupled-loop system using optical amplification and Mach-Zehnder modulation. In magnonic nanowaveguides, Merbouche et al.~\cite{NC2024Merbouche} used pulsed spin-orbit torque to control effective damping, accessing both coherent attenuation and amplification within a finite time window.

On the other hand, the available control parameters in a specific platform might be not sufficient to activate all the biorthogonal passages and simultaneously restore their standard norms at the final time as we have done in Secs.~\ref{sec:popu_transfer} and \ref{sec:gate}, i.e., the coherent control over the whole system is not always guaranteed. Nevertheless, we can relax the full-rank control by focusing merely on a few selected passages spanning a desired subspace, in which an effective unitary transformation can be still available in an open quantum system.

In summary, we have developed a biorthogonal quantum control framework for finite-dimensional systems governed by the time-dependent non-Hermitian Hamiltonian, targeting on universal quantum control in biorthogonal nonadiabatic passages. With similarity transformation, the prescribed ancillary basis defines a generally nonunitary instantaneous frame with a non-Hermitian gauge potential. We have proved that the von Neumann equation for the desired ancillary projector constructed by both ket and bra passages is necessary and sufficient to activate the relevant pair of dual passages. When this condition holds for a complete set of biorthogonal bases, the rotated Hamiltonian becomes diagonal in stationary frame, allowing arbitrary initial states to evolve as superpositions of transitionless passages. Extra conditions on the cancellation of the imaginary parts of the accumulated passage phases render a unitary endpoint transformation without artificial state normalization.

Our theory is justified in the two-level and three-level systems for population transfer and holonomic gate, respectively. The results establish a solid connection between a complete biorthogonal nonadiabatic control and coherent quantum operations during the nonunitary dynamics. By relieving the endpoint unitary condition, we provide a complementary viewpoint that reveals key features of chiral state conversion in a non-Hermitian two-level system with and with no EP encircling. In our framework, this conversion is induced by the chiral-dependent complex phases accumulated along the loops. Reversing the looping direction in the parametric space changes the signs of the imaginary phases, thereby producing different final eigenstates.

\section*{Acknowledgments}

We acknowledge grant support from the National Natural Science Foundation of China (Grant No. U25A20199) and the ``Pioneer'' and ``Leading Goose'' R\&D of Zhejiang Province (Grant No. 2025C01028).

\bibliographystyle{apsrevlong}
\bibliography{bior_uqc_ref}

\end{document}